\documentclass[aps,prd,twocolumn,eqsecnum]{revtex4}
\begin{document}
\title{The Tolman-Ehrenfest effect for an ideal gas in a  background of time-independent electric, magnetic, and gravitational fields }
\author{H. Arthur Weldon}
\email[]{hweldon@WVU.edu}
\affiliation{Department of Physics and Astronomy, West Virginia University , Morgantown, West Virginia, 26506-6315}
\date{\today}

\begin{abstract} The statistical mechanics of an ideal gas of point particles moving in a 
  time independent background metric with $g_{0j}\neq 0$ is investigated. An explicit calculation shows  that  when there
 is no background electrostatic or magnetostatic field the thermodynamic pressure, energy density, and  thermally averaged
energy-momentum tensor  depend on temperature and chemical potential
only through the ratios $T_{0}/\sqrt{g_{00}}$ and $\mu_{0}/\sqrt{g_{00}}$. A  background magnetostatic field does
not change this, 
however with a  background electrostatic field the previous results are multiplied by a factor $\exp(-eA_{0}/T_{0})$, 
which is an exception to the strict Tolman-Ehrenfest rule because the system is open.
\end{abstract}
\maketitle

      \section{Introduction} 
The Tolman-Ehrenfest  effect \cite{Tolman1,Tolman2,Tolman3} originated with the observation that classical electromagnetic radiation in thermal equilibrium
with matter in a static background metric ($\partial_{0}g_{\mu\nu}=0$ and  $g_{0j}=0$) would have a thermally averaged energy-momentum
tensor   whose temperature dependence always occurs in the ratio
 $T_{0}/\sqrt{g_{00}}$, where $T_{0}$ is spacetime independent. 
 There are a number of arguments \cite{Sanchez,Lima,Cremaschini,Kim,
Faraoni,Kovtun,Munoz} that support the Tolman-Ehrenfest result for the static case $g_{0j}=0$.
Ref. \cite{Munoz} specifically treats the ideal gas. 

There are   arguments that the Tolman-Ehrenfest effects is valid when $g_{0j}\neq 0$
 \cite{Rovelli1,Haggard,Rovelli2,Wald,Santiago1,Santiago2,Xia,Santiago3,Chernodub}.
A special example in this category is that of a rotating Minkowski reference frame \cite{Santiago3,Chernodub} in which
$g_{t\phi}$ is non-zero but there is no curvature.

 The Schwarzschild metric illustrates the issue.  In the original $(t,r,\theta,\phi)$  coordinates the metric is static:
 \begin{equation}
 (ds)^{2}=\Big(1-{2GM\over r}\Big)(dt)^{2}-{(dr)^{2}\over 1-2GM/r}-r^{2}d\Omega^{2}
 \end{equation}
 with $d\Omega^{2}=d\theta^{2}+\sin^{2}\theta\, d\phi^{2}$. The thermodynamic functions for an ideal gas in this
 background metric will depend on $T_{0}$ only through the ratio $T_{0}/\sqrt{1-2GM/r}$.
 A change to the outgoing Eddington-Finkelstein time coordinate \cite{Eddington,Finkelstein}
 \begin{equation}t'=t-2GM\ln(r-2GM)\end{equation}
 changes the form of the line element to
 \begin{eqnarray}
 (ds)^{2}&=&\Big(1-{2GM\over r}\Big)(dt')^{2}+{4GM\over r} dt'dr\nonumber\\
 &-&\Big(1+{2GM\over r}\Big)(dr)^{2}-r^{2}d\Omega^{2}.
 \end{eqnarray}
 Since $g_{t'r}\neq 0$ this metric is not static but stationary.  It will follow from the calculation
 described herein that
 an ideal gas of point particles in this metric will be the same function of $T_{0}/\sqrt{1-2GM/r}$.

 The more important situations are those in which the metric is stationary
 and cannot be changed to static by a coordinate transformation. 
  Though in  this paper the metric is not required to satisfy the Einstein field equations, there are two
familiar stationary metrics that do solve the field equations and are  not coordinate equivalent
to a static metric:
  the  Kerr metric  describing an uncharged, but rotating black hole; and the Kerr-Newmann metric
  describing a charged, rotating black hole \cite{Newman}.  
  In Boyer-Lindquist coordinates $(t,r,\theta,\phi)$ both
  metrics have $g_{t\phi}\neq 0$.
  The Kerr-Newman black hole is surrounded by a static electric field and a
  static magnetic field.  There are many other stationary metrics that solve the field equations  \cite{Stephani}.
  A limitation of the present analysis is that equilibrium
  statistical mechanics, whether done in the canonical or the grand canonical ensemble, applies to
  situations in which the total number of particles is conserved. Thus if the metric has an event singularity
   statistical mechanics is only applicable well outside the event horizon.

 The Kerr-Newman example motivates extending the investigation to include
 arbitrary background electrostatic and/or magnetostatic fields
and this leads to a specific exception to the
 Tolman-Ehrenfest effect. The exception  can be illustrated by a simple example.
 Consider an ideal gas of nonrelativistic point particles with no gravitational field.
In kinetic theory the distribution function for such a gas in thermal equilibrium is
\begin{equation}
f_{0}=n_{0}\Big({2\pi\over mT_{0}}\Big)^{3/2}e^{-{\bf p}^{2}/2mT_{0}}.
\end{equation}
The particle density $n_{0}=\int [d^{3}p/(2\pi)^{3}] f_{0}$ is spatially uniform.
Suppose the particles have charge $e$ and a time-independent external
electric and magnetic field is imposed.  After equilibrium is achieved the new distribution function
should satisfy the Vlasov equation  \cite{Landau2}
\begin{equation}
{\partial f\over\partial t}+v^{j}{\partial f\over\partial x^{j}}
+e({\bf E}+{\bf v}\times{\bf B})_{j}{\partial f\over\partial p_{j}}=0.
\end{equation}
In a complete analysis ${\bf E}$ and ${\bf B}$  would be the sum of the external fields
and the internal fields; the internal fields being determined self-consistently by solving 
Maxwell's  equations with the charge density and current density computed from the distribution
function $f$. 
If the external fields are much stronger than the internal fields produced by the charged particles
then ${\bf E}$ and ${\bf B}$ may taken as the external fields, with time-independent
scalar and vector potentials $A_{0}$ and  ${\bf A}$.
The solution to the Vlasov equations is 
\begin{equation}
f({\bf x},{\bf p})=n_{0}\Big({2\pi\over mT_{0}}\Big)^{3/2}e^{-[({\bf p}-e{\bf A})^{2}/2m+eA_{0}]/T_{0} }
\end{equation}
where $A_{0}$ and ${\bf A}$ are evaluated at the position ${\bf x}(t)$ of the particle  and $f$ has no explicit time dependence. 
The particle density is 
\begin{equation}
n({\bf x})=\int {d^{3}p\over (2\pi)^{3}}f({\bf x},{\bf p})=n_{0}e^{-eA_{0}({\bf x})/T_{0}}.
\end{equation}
The factor $\exp[-eA_{0}/T_{0}]$ will appear in  other  thermodynamic functions:
pressure, energy density, entropy density.
(The  gauge transformation that eliminates $A_{0}$ would make ${\bf A}$ time-dependent and
the distribution function would no longer satisfy the Vlasov equation.)
It will turn out that in the presence of a stationary gravitational field the same factor $\exp[-eA_{0}/T_{0}]$
will occur,  whereas the Tolman-Ehrenfest expectation would be a
factor  $\exp[-eA_{0}\sqrt{g_{00}}/T_{0}]$.  This is not in conflict with derivations of the Tolman-Ehrenfest
effect \cite{Sanchez,Lima,Cremaschini,Kim,
Faraoni,Kovtun,Munoz,Rovelli1,Haggard,Rovelli2,Wald,Santiago1,Santiago2,Xia,Santiago3} which rely on
$(T^{\mu\nu})_{;\mu}=0$.

Throughout the discussion the particles are thermalized; the internal electric and magnetic fields
produced by the charged particles are neglected and $A_{0}$, ${\bf A}$ are  unthermalized,
external potentials.

The central problem is that the non-vanishing of $g_{0\ell}$ makes the Hamiltonian 
for a point particle moving on the geodesics of 
 a time-independent background metric rather complicated. It will be shown in Sec. II that the particle Hamiltonian 
 in the absence of a background electrostatic or magnetostatic field is
\begin{equation}
\overline{H}={1\over g^{00}}\bigg[\sqrt{g^{00}(m^{2}\!-\!g^{ij}p_{i}p_{j}) \!+\!(g^{0\ell}p_{\ell})^{2}}+ g^{0\ell}p_{\ell}\bigg].
\end{equation}
Here $p_{j}(t)$ are the canonical momenta and the metric  components are evaluated at the particle
position  $x^{i}(t)$.
Though covariant metric components $g_{\alpha\beta}$ do not appear in $\overline{H}$, the minimum value of $\overline{H}$ occurs at $p_{j}=-mg_{j0}/\sqrt{g_{00}}$ and the minimum energy
is $m\sqrt{g_{00}}$. 

\paragraph*{Outline.} 
 Sec. II derives the Hamiltonian $\overline{H}$ and incorporates   a  time-independent 
background  of static electric and magnetic fields. The Hamiltonian leads to
 the partition function and the thermodynamic pressure in the grand canonical ensemble
in terms of two parameters $T_{0}$ and a chemical potential $\mu_{0}$.
The pressure due to the particles is of the form $e^{\beta_{0}(\mu_{0}-eA_{0})}\overline{P}$.

Sec. III computes  $\overline{P}$ explicitly and after using a particular addition theorem for
Bessel functions the final result is 
\begin{equation}
\overline{P}={m^{2}\over 2\pi^{2}}\Big({T_{0}\over \sqrt{g_{00}}}\Big)^{2}
K_{2}\Big(m{\sqrt{g_{00}}\over T_{0}}\Big)
\end{equation}
 with no dependence on $g_{0\ell}$ or $g_{jk}$. 

Sec. IV contains the computation of the thermally averaged energy-momentum tensor. 
The particle contribution  has the perfect fluid form
\begin{equation}
\langle T^{\mu\nu}_{\rm part}\rangle=e^{\beta_{0}(\mu_{0}-eA_{0})}
\big[U^{\mu}U^{\nu}(\overline{\rho}+\overline{P})-g^{\mu\nu}\overline{P}\big].
\end{equation}

Sec. V discusses  the low temperature limit 
and how to change from classical Boltzman statistics to Bose or Fermi statistics. 

Appendix A details the change of integration variables from the canonical momenta to 
Euclidean momenta that makes possible the integration in Sec. III.

Appendix   B proves a result used in Sec. IV, namely that the thermally averaged energy-momentum tensor may be calculated from
the variational derivative of the partition function with respect to the time-independent metric:
\begin{equation}
{\delta\ln Z\over\delta g_{\mu\nu}}=
-{\beta_{0}\sqrt{g}\over 2}\Big[\langle T^{\mu\nu}_{\rm part}\rangle +T^{\mu\nu}_{\rm field}\Big].
\end{equation}
Because $g_{0j}\neq 0$ it is not trivial to compute ${\cal H}_{\rm field}$  in terms of the canonical momenta $\pi^{j}$ and the covariant fields $F_{jk}$. The result is displayed in (\ref{263}). The derivative of this Hamiltonian density
with respect to the metric, keeping $\pi^{j}$  and $F_{jk}$ fixed,  yields the usual Hilbert energy-momentum
tensor $T^{\mu\nu}_{\rm field}$ when the momenta are re-expressed in terms of field strengths. 

Appendix C gives specific results for the energy density, number density, and entropy density. At low temperature
the entropy density is given by a local form of the Sackur-Tetrode equation.   

Greek letters run over 0,1,2,3; Latin letters over 1,2,3.
The metric signature is $(+---)$ and $g=|{\rm det}(g_{\mu\nu})|$;
and $\hbar=c=1$.

                 \section{Hamiltonian and partition function} The first step is to derive the  expression  (\ref{268})  for the 
 thermodynamic pressure. The starting point is the   action for $N$ particles in a background comprised of a time-independent metric 
plus  electrostatic and magnetostatic fields \cite{Weinberg}
\begin{eqnarray}
S&=&-\sum_{n=1}^{N}\Big\{m\int \sqrt{g_{\mu\nu}(x_{n})dx_{n}^{\mu}dx_{n}^{\nu}}
+e\int A_{\mu}(x_{n})dx_{n}^{\mu}\Big\}\nonumber\\
&&-\int d^{4}x{\sqrt{g}\over 4}  F_{\mu\nu}(x)F^{\mu\nu}(x).
\end{eqnarray}
The particle coordinates $x_{n}^{j}$ depend on the time coordinate $x_{n}^{0}$ or equivalently
on the proper time $d\tau_{n}=g_{\mu\nu}(x_{n})dx_{n}^{\mu}dx_{n}^{\nu}$.  With  $g_{0j}\neq 0$ the
proper time is not time reversal invariant and the particle motion is not time-reversal invariance. 

\paragraph{Hamiltonian for N particles.}
The Lagrangian for $N$ particles is a sum of identical terms
\begin{displaymath}
L_{N}=\sum_{n=1}^{N} L(x_{n}^{i},v^{j}_{n})\end{displaymath}
where $v_{n}^{j}=dx^{j}/dx^{0}_{n}$ and 
each term has the form
\begin{displaymath}
L(x^{i},v^{j})=-m\sqrt{g_{00}+2g_{j0}v^{j}+g_{jk} v^{j}v^{k}}-e(A_{0}+A_{j}v^{j}).
\end{displaymath}
The metric and the vector potential are evaluated at the position of the particle $x^{k}(t)$.
The canonical momentum is
\begin{equation}
p_{j}={\partial L\over\partial v^{j} }={-m(g_{j0}+g_{jk}v^{k})\over
\sqrt{g_{00}+2g_{0i}v^{i}+g_{i\ell}v^{i}v^{\ell}}}
-eA_{j}\label{229}.\end{equation}
In terms of the velocity $v^{j}=dx^{j}/dt$ this means that
\begin{eqnarray}
H_{\rm part}&=&p_{j}v^{j}-L\nonumber\\
&=&{m(g_{00}+g_{0j}v^{j})\over \sqrt{g_{00}+2g_{0i}v^{i}+g_{i\ell}v^{i}v^{\ell}}}+eA_{0}.\label{230}
\end{eqnarray}
To express the velocity in terms of the momentum it is convenient to define
\begin{equation}
p_{j}^{\prime}=p_{j}+eA_{j}.\label{232}\end{equation}
Note that $p_{j}$ depends only on $t$ and $A_{j}$ depends on $x^{i}(t)$.
Eq.  (\ref{229}) may be inverted to express the velocity in terms
of momenta
\begin{equation}
v^{j}={g^{j0}\over g^{00}}-{c^{jk}p^{\prime}_{k}\over\sqrt{g^{00}[m^{2}-c^{i\ell}
p^{\prime}_{i}p^{\prime}_{\ell}]}}.\label{234}\end{equation}
The matrix  $c^{jk}$ is given by
\begin{equation}
c^{jk}=g^{jk}-{g^{j0}g^{k0}\over g^{00}}\label{236}\end{equation}
and satisfies
\begin{equation}
g_{ij}c^{jk}=\delta_{i}^{k}.\label{240}
\end{equation}
(For the special case in which $g_{ij}$ is diagonal, one can calculate $g^{\mu\nu}$ by Cramer's rule and show that
$c^{jk}$ is diagonal and $c^{jj}=1/g_{jj}$. )
The Hamiltonian for each particle is 
\begin{eqnarray}
&&H_{\rm part}=\overline{H}+eA_{0}\nonumber\\
&&\overline{H}={1\over g^{00}}\bigg[\sqrt{g^{00}(m^{2}\!-\!c^{i\ell}p^{\prime}_{i}p^{\prime}_{\ell})}+g^{0\ell}p^{\prime}_{\ell}\bigg]
.\label{245}\end{eqnarray}
The minimum of $\overline{H}$ is $m\sqrt{g_{00}}$ and occurs at
 $p^{\prime}_{j}=-mg_{j0}/\sqrt{g_{00}}$.
For small velocity $\overline{H}$ may be expanded as
\begin{equation}
\overline{H}=m\sqrt{g_{00}}\bigg\{1-{g_{ij}v^{i}v^{j}\over 2g_{00}}+2\Big[{g_{0\ell}v^{\ell}\over g_{00}}\Big]^{2}
+{\cal O}(v^{3})\bigg\}
\end{equation}
which shows that for small velocity if  $g_{0\ell}v^{\ell}\neq 0$ the energy is higher than for a static metric.

 \paragraph{The equation of motion.} Though equilibrium statistical mechanics does not employ the
 equation of motion it is important to check that the Hamiltonian does give the correct equation of
 motion for the particles. 
 Hamilton's first equation
 \begin{equation}
 v^{j}={\partial H_{\rm part}\over\partial p_{j}}\label{250}\end{equation}
 reproduces  (\ref{234}).  
 Hamilton's second equation 
  \begin{equation}
 -{dp_{j}\over dt}={\partial H_{\rm part}\over\partial x^{j}}\label{252}
 \end{equation}
 will yield the  equation of motion.  It is convenient to employ
 the proper velocity 
 \begin{equation}
 u^{\mu}={dx^{\mu}\over d\tau}\label{253}
 \end{equation}
with components
 \begin{eqnarray}
 u^{j}&=&{v^{j}\over  \sqrt{g_{00}+2g_{0i}v^{i}+g_{i\ell}v^{i}v^{\ell}}}\label{254a}\\
 u^{0}&=&{1\over  \sqrt{g_{00}+2g_{0i}v^{i}+g_{i\ell}v^{i}v^{\ell}}}.\label{254b}
 \end{eqnarray}
 Since $u_{\alpha}=g_{\alpha\mu}u^{\mu}$, reference to (\ref{229}) and (\ref{230})
 shows that
 \begin{eqnarray}
 p_{j}&=&-mu_{j}-eA_{j}\label{255a}\\
 H_{\rm part}&=&mu_{0}+eA_{0}.\label{255b}
 \end{eqnarray}
 Hamilton's second equation (\ref{252})
 becomes
 \begin{equation}
 m{du_{j}\over dt}+e{dA_{j}\over dt}=m{\partial u_{0}\over \partial x^{j}}+e{\partial A_{0}\over \partial x^{j}}
. \label{256}\end{equation}
 A simple way to compute the spatial derivative of $u_{0}$ is to apply
  $\partial /\partial x^{j}$ to $m^{2}=g^{\mu\nu}u_{\mu}u_{\nu}$, which results in
 \begin{equation}
 u^{0}{\partial u_{0}\over\partial x^{j}}=-{1\over 2}(\partial_{j} g^{\mu\nu})u_{\mu}u_{\nu}+eu^{k}\partial_{j}A_{k}.
 \end{equation}
 Multiplying (\ref{256}) by $u^{0}$ and using $u^{0}(d/dt)=d/d\tau$ gives
 \begin{equation}
 m\Big[{du_{j}\over d\tau}\!-\!{1\over 2}(\partial_{j}g_{\mu\nu})u^{\mu}u^{\nu}\Big]=eF_{j\alpha}u^{\alpha}.\label{258}
 \end{equation}
  As yet there are only three equations. To obtain the fourth
contract  (\ref{258}) with $u^{j}$ and use $u^{j}(du_{j}/d\tau)=
 -u^{0}(du_{0}/d\tau)$ to obtain
 \begin{equation}
 m{du_{0}\over d\tau}=eF_{0j}u^{j}.\label{259}\end{equation}
 The four components of  (\ref{258}) and (\ref{259}) are summarized by
 \begin{equation}
 m\Big({du_{\lambda}\over d\tau}-{1\over 2}(\partial_{\lambda}g_{\mu\nu})u^{\mu}u^{\nu}\Big)=eF_{\lambda\alpha}u^{\alpha}
 .\label{260}\end{equation}
The contravariant form of this equation is
 \begin{equation}
 m\Big({du^{\lambda}\over d\tau}+\Gamma^{\lambda}_{\mu\nu}g_{\mu\nu}u^{\mu}u^{\nu}\Big)=eF^{\lambda\alpha}u_{\alpha}
 ,\label{262}\end{equation}
 which is the correct equation of motion for a particle in a background electric/magnetic field \cite{Landau1}.
 Solutions  for the particle trajectories are obtained in \cite{Hackmann}.
 
\paragraph{Hamiltonian for the background electric/magnetic field.} The Hamiltonian for the background electromagnetic field is also needed.
The canonical Hamiltonian density given by Noether's theorem
 in terms of the canonical momenta $\pi^{j}=\partial {\cal L}_{\rm field}/\partial(\partial_{0}A_{j})$ and canonical fields $F_{jk}$ is 
\begin{eqnarray}
{\cal H}_{\rm field}&=&-{\pi^{j}\pi^{k}g_{jk}\over 2g^{00}\sqrt{g}}+{\pi^{j}F_{jk}g^{k0}\over g^{00}}\nonumber\\
&+&{\sqrt{g}\over 4}F_{jk}F_{\ell m}c^{j\ell}c^{km}+\pi^{j}\partial_{j}A_{0}\label{263}
\end{eqnarray}
as shown in Appendix C.

\paragraph{Partition function for the ideal gas.} 
 In a gas of $N$ particles the Hamiltonian for each particle
is of the form (\ref{245}) and is a function of each particle's contravariant position coordinates $x^{1},x^{2},x^{3}$  and its covariant
momenta $p_{1},p_{2},p_{3}$.  The Hamiltonian for $N$ particles is the sum of the single particle Hamiltonians:
\begin{equation}
H_{N}=\sum_{n=1}^{N}H_{\rm part}(x^{i}_{n}, p^{n}_{j})\end{equation}
and the total Hamiltonian is $H_{N}+\int d^{3}x{\cal H}_{\rm field}$.
The Boltzman factor $e^{-\beta_{0}H_{N}}$ is a product of N exponentials.
The  integration of each factor $e^{-\beta_{0}H_{\rm part}}$  over its six-dimensional phase space is the same and so  the  partition function for N particles  is 
\begin{equation}
Z_{N}={1\over N!} \Big[ \int {d^{3}x d^{3}p\over (2\pi)^{3} }e^{-\beta_{0}H_{\rm part}}\Big]^{N}e^{-\beta_{0}H_{\rm field}}
.\end{equation}
(Because of  the convention $\hbar\to 1$ the denominator  $h^{3}=(2\pi\hbar)^{3}\to (2\pi)^{3}$.)

In the grand canonical ensemble
the partition function depends on the chemical potential $\mu_{0}$:
\begin{equation}
Z=\sum_{N=0}^{\infty}(e^{\beta_{0}\mu_{0}})^{N}Z_{N},\end{equation}
and therefore
\begin{equation}
\ln Z=\int {d^{3}x d^{3}p\over (2\pi)^{3} }e^{-\beta_{0}(H_{\rm part}-\mu_{0})}-\beta_{0}H_{\rm field}.
\end{equation}
The momentum integration variable may be shifted from the canonical $p_{j}$ to $p_{j}^{\prime}$
\begin{equation}
p_{j}^{\prime}=p_{j}+eA_{j};\label{264}\end{equation}   this removes
$A_{j}$ from the partition function as expected from the Bohr-Van Leeuwen theorem
\cite{Huang}.

The partition function is directly related to the thermodynamic pressure \cite{Dowker}
\begin{equation}
\ln Z=\beta_{0}\int d^{3}x\sqrt{g}\,P\label{265}.\end{equation}
Therefore
\begin{equation}
 P=P_{\rm part} -{1\over\sqrt{g}}{\cal H}_{\rm field},\label{266}\end{equation}
where the  pressure exerted by the gas of particles is
\begin{equation}
P_{\rm part}={T_{0}\over\sqrt{g}}
\int {d^{3}p^{\prime}\over(2\pi)^{3}}e^{-\beta_{0}(H_{\rm part}-\mu_{0})}.\label{268}
\end{equation}
The next step is to perform this integration.
 
      \section{Calculation of the particle pressure} 
The terms in $H_{\rm part}$  involving $\mu_{0}$ and  $eA_{0}$ have no momentum dependence and so
\begin{eqnarray}
P_{\rm part}&=&e^{\beta_{0}(\mu_{0}-eA_{0})} \overline{P}\label{302}\\
\overline{P}&=&{T_{0}\over\sqrt{g}}
\int {d^{3}p'\over(2\pi)^{3}}e^{-\beta_{0}\overline{H}}.\label{304}
\end{eqnarray}
The dependence on $\mu_{0}$ may be written
\begin{equation}
\beta_{0}\mu_{0}={\sqrt{g_{00}}\over T_{0}}{\mu_{0}\over\sqrt{g_{00}}}.\label{305}\end{equation}
and is in agreement with Oscar Klein's argument \cite{Klein} that both $T_{0}$ and  $\mu_{0}$ will always occur
divided by $\sqrt{g_{00}}$.
The factor $\exp[-e A_{0}/T_{0}]$ in (\ref{302})
is an exception to the Tolman-Ehrenfest rule and will remain in the final result. 
This section will show that $\overline{P}$ is given by (\ref{391})
and is a function only of the ratio $T_{0}/\sqrt{g_{00}}$.

\paragraph{Euclidean momenta.}   To compute $\overline{P}$ it is necessary
  to change the integration variables from 
  $p_{j}^{\prime}$ with metric $g_{jk}$ into Euclidean momenta $k_{a}$. The details of this transformation are given in Appendix A.
The Hamiltonian becomes 
\begin{equation}
\overline{H}={1\over\sqrt{g^{00}}}\Big[\sqrt{m^{2}+{\bf k}^{2}}+{\bf s}\cdot{\bf k}\Big]\label{306},\end{equation}
where ${\bf s}$ is a Euclidean vector with length squared
\begin{equation}
{\bf s}^{2}=g_{0j}g^{j0}=1-g_{00}g^{00}.\label{308}\end{equation}
The minimum of $\overline{H}$ occurs at $k_{a}=-m{\rm s}_{a}\big/\sqrt{1-{\rm s}^{2}}$
and the value of the minimum is still $m\sqrt{g_{00}}$. The change in integration variables requires
\begin{equation}
d^{3}p^{\prime}=\sqrt{gg^{00}}\, d^{3}k,\label{310}\end{equation}
which leads to
\begin{equation}
\overline{P}=T_{0}\sqrt{g^{00}}
\int {d^{3}k\over(2\pi)^{3}}e^{-\beta_{0}\overline{H}}.\label{312}
\end{equation}

\paragraph{Dimensionless variables.} The change to a dimensionless
integration variable ${\bf u}={\bf k}/m$
and introduction of a dimensionless parameter
\begin{equation}
z={m\over T_{0}\sqrt{g^{00}}}\label{320}.\end{equation}
converts the  exponent  in the integrand of (\ref{304}) to
\begin{equation}
\beta_{0}\overline{H}=z\big[\sqrt{1+{\bf u}^{2}}+{\bf s}\cdot{\bf u}\big].\label{325}\end{equation}
The pressure integral is
\begin{equation}
\overline{P}(s,z)={m^{4}\over z}\int {d^{3}u\over (2\pi)^{3}}
e^{-\beta_{0}\overline{H}}.\label{335}
\end{equation}
 It  appears that  $\overline{P}$   is a function of the two variables $s$ and $z$. To show that it actually depends only
on a particular combination of these two variables requires performing  the integration.
The angular integration gives
\begin{equation}
\overline{P}(s,z)={m^{4}\over 2\pi^{2}s z^{2}}\int_{0}^{\infty}\!\! udu\,e^{-z\sqrt{1+u^{2}}}
\sinh[szu].\end{equation}
Next  expand the $\sinh$ in an infinite series:
\begin{equation}
\overline{P}(s,z)={m^{4}\over 2\pi^{2}z}\sum_{\ell=0}^{\infty}{(sz)^{2\ell}\over (2\ell\!+\!1)!}
\int_{0}^{\infty}du\,e^{-z\sqrt{1+u^{2}}}u^{2\ell+2}.
\end{equation}
The necessary integrals are modified Bessel functions of the second kind \cite{Gradshteyn}:
\begin{equation}
\int_{0}^{\infty}du\,e^{-z\sqrt{1+u^{2}}}u^{2\ell+2}={\Gamma(\ell+{3\over 2})\over\Gamma({1\over 2})}
\Big({2\over z}\Big)^{\ell+1}K_{\ell+2}(z)
\end{equation}
and the pressure becomes
\begin{equation}
\overline{P}(s,z)={m^{4}\over 8\pi^{2}}\sum_{\ell=0}^{\infty}{s^{2\ell}\over \ell!}\Big({z\over 2}\Big)^{\ell-2}K_{\ell+2}(z).\label{350}
\end{equation}

By using the asymptotic value of $K_{\ell+2}(z)$ either for low temperature  ($z\gg 1$)
or for high  temperature  ($z\ll 1$) one can perform the sum on $\ell$ and obtain the leading term and the sub-leading term
in either regime and confirm the Tolman-Ehrenfest effect to that order. This suggests that the infinite
series representation can be simplified.

\paragraph{Simpler expression for $\overline{P}$.}
It will turn out that the entire series is only a function of
the single variable
\begin{equation}
z\sqrt{1-s^{2}}=m{\sqrt{g_{00}}\over T_{0}}.\label{380}
\end{equation}
To show this  define the derivative combination
\begin{equation}
{\cal D}=z{\partial\over\partial z}+({1\over s}-s){\partial\over\partial s},
\end{equation}
with the property 
\begin{equation}
{\cal D}(z\sqrt{1-s^{2}})=0.\end{equation}
Application of ${\cal D}$ to  (\ref{350}) gives
\begin{eqnarray}
{\cal D}\overline{P}(s,z)&=&{m^{4}\over 8\pi^{2}}\sum_{\ell=0}^{\infty}\bigg[{s^{2\ell}\over\ell!}2\ell({1\over s^{2}}-1)\Big({z\over 2}\Big)^{\ell-2}K_{\ell+2}(z)\nonumber\\
&&+{s^{2\ell}\over\ell!}(\ell-2)\Big({z\over 2}\Big)^{\ell-2}K_{\ell+2}(z)\\
&&+{s^{2\ell}\over\ell!}\Big({z\over 2}\Big)^{\ell-2}z{d\over dz}K_{\ell+2}(z)\bigg].\nonumber
\end{eqnarray}
The last line is simplified by using the identity \cite{Gradshteyn}
\begin{equation}
z{d\over dz}K_{\ell+2}(z)=(\ell+2)K_{\ell+2}(z)-zK_{\ell+3}(z).\label{386}
\end{equation}
All  terms proportional to $K_{\ell+2}$ combine:
\begin{eqnarray}
{\cal D}\overline{P}(s,z)&=&{m^{4}\over 8\pi^{2}}\sum_{\ell=1}^{\infty}{s^{2\ell-2}\over(\ell\!-\!1)!}2 \Big({z\over 2}\Big)^{\ell-2}K_{\ell+2}(z)\nonumber\\
&-&{m^{4}\over 8\pi^{2}}\sum_{\ell=0}^{\infty}{s^{2\ell}\over\ell!}\Big({z\over 2}\Big)^{\ell-2} z K_{\ell+3}(z).
\end{eqnarray}
The two series cancel and thus  ${\cal D}\overline{P}(s,z)=0$ and so  $\overline{P}(s,z)$ is a function only of the single variable (\ref{380}):
\begin{equation}
\overline{P}(s,z)={m^{4}\over 8\pi^{2}}\Psi(z\sqrt{1-s^{2}}\,),\end{equation} 
or equivalently
\begin{equation}
\sum_{\ell=0}^{\infty}{s^{2\ell}\over \ell!}\Big({z\over 2}\Big)^{\ell-2}K_{\ell+2}(z)
=\Psi(z\sqrt{1-s^{2}}).
\end{equation}
To obtain an explicit form for $\Psi$, set $s=0$ 
\begin{equation}
\Big({2\over z}\Big)^{2}K_{2}(z)=\Psi(z).
\end{equation}
This determines the function $\Psi$ and so for $s\neq 0$ 
\begin{equation}
\overline{P}(s,z)={m^{4}\over 2\pi^{2}} {K_{2}(z\sqrt{1-s^{2}}\,)\over [z\sqrt{1-s^{2}}]^{2}}.\label{390}\end{equation}
In more physical variables
\begin{equation}
\overline{P}={m^{2}\over 2\pi^{2}}\Big({T_{0}\over \sqrt{g_{00}}}\Big)^{2}
K_{2}\Big(m{\sqrt{g_{00}}\over T_{0}}\Big).\label{391}
\end{equation}
(That the two expressions (\ref{350}) and (\ref{391}) for $\overline{P}$ are equal amounts to
\begin{equation}
{1\over 4}\sum_{\ell=0}^{\infty}{s^{2\ell}\over \ell!}\Big({z\over 2}\Big)^{\ell-2}K_{\ell+2}(z)
={K_{2}(z\sqrt{1-s^{2}}\,)\over [z\sqrt{1-s^{2}}]^{2}}\label{392},\end{equation}
which is a known addition theorem for Bessel functions \cite{Watson}.)

\paragraph{Energy density.}  For later purposes it is convenient to
compute the quantity 
\begin{equation}
\overline{\rho}={T_{0}\over\sqrt{g}}\int {d^{3}p'\over(2\pi)^{3}}\overline{H}e^{-\beta_{0}\overline{H}}
=T_{0}{\partial\overline{P}\over\partial T_{0}}-\overline{P}\label{394}.\end{equation}
Using (\ref{386}) gives
\begin{equation}
\overline{\rho}+\overline{P}=
{m^{3}\over 2\pi^{2}}\Big({T_{0}\over \sqrt{g_{00}}}\Big)
K_{3}\Big(m {\sqrt{g_{00}}\over T_{0}} \Big).\label{395}
\end{equation}

       \section{The energy-momentum tensor}
Appendix B  shows that the thermally averaged energy-momentum tensor may be computed
from the variation derivative of the partition function. Eqs. (\ref{B20}) and (\ref{B90}) give 
\begin{equation}
{\delta\ln Z\over\delta g_{\mu\nu}}=
-{\beta_{0}\sqrt{g}\over 2}\Big[\langle T^{\mu\nu}_{\rm part}\rangle +T^{\mu\nu}_{\rm field}\Big].\label{400}
\end{equation}
Since $\ln Z=\beta_{0}\int d^{3}x\sqrt{g}P$ this is equivalent to
\begin{equation}
\langle T^{\mu\nu}_{\rm part}\rangle+T^{\mu\nu}_{\rm field}=-{2\over\sqrt{g}}{\partial(\sqrt{g}P)\over\partial g_{\mu\nu}}
.\end{equation}
 Using $\sqrt{g}P$ from (\ref{266}) gives
\begin{eqnarray}
\langle T^{\mu\nu}_{\rm part}\rangle&=&-{2\over\sqrt{g}}{\partial(\sqrt{g}P_{\rm part})\over\partial g_{\mu\nu}}\label{410}\\
T^{\mu\nu}_{\rm field}&=&{2\over\sqrt{g}} {\partial {\cal H}_{\rm field}\over\partial g_{\mu\nu}}\label{412}
.\end{eqnarray}

\paragraph{Field EMT.}
In Appendix B, ${\cal H}_{\rm field}$ is expressed in terms of canonical momenta $\pi^{j}$ and canonical
fields $F_{jk}$ with the result (\ref{B80}). The partial derivative (\ref{412}) is computed keeping
$\pi^{j}$ and $F_{jk}$ fixed; when the result is re-expressed  in terms of the fields the result is
the usual Hilbert energy-momentum tensor:
\begin{equation}
T^{\mu\nu}_{\rm field}=-F^{\mu\alpha}F^{\nu\beta}g_{\alpha\beta}+{g^{\mu\nu}\over 4}F_{\alpha\beta}F^{\alpha\beta}
.\label{415}\end{equation}

\paragraph{Particle EMT.}
To implement  Eq. (\ref{410}) begin with
$P_{\rm part}=e^{\beta_{0}(\mu_{0}-eA_{0})}\overline{P}$.
Since  $\overline{P}$ depends only on the $g_{00}$ component of the metric the derivative in (\ref{410}) is
\begin{equation}
\langle T^{\mu\nu}_{\rm part}\rangle=e^{\beta_{0}(\mu_{0}-eA_{0})}
\Big[-2\delta^{\mu}_{0}\delta^{\nu}_{0}{\partial\overline{P}\over\partial g_{00}}-g^{\mu\nu}\overline{P}\Big]\end{equation}
As shown in  (\ref{390})  the $g_{00}$ dependence of $\overline{P}$ occurs  only through  the ratio $T_{0}/\sqrt{g_{00}}$ 
and so a $g_{00}$ derivative is equivalent to a $T_{0}$ derivative:
\begin{equation}
-2g_{00}{\partial\overline{P}\over\partial g_{00}}=T_{0}{\partial\overline{P}\over\partial T_{0}}
=\overline{\rho}+\overline{P}.
\end{equation}
Therefore
\begin{equation}
\langle T^{\mu\nu}_{\rm part}\rangle=e^{\beta_{0}(\mu_{0}-eA_{0})}
\Big\{ \delta^{\mu}_{0}\delta^{\nu}_{0}{(\overline{\rho}+\overline{P})\over g_{00}}-g^{\mu\nu}\overline{P}\Big\}
.\label{420}\end{equation}
The thermal average of the particle velocity is zero because $v^{j}=\partial H_{\rm part}/\partial p_{j}$: 
\begin{equation}
\int {d^{3}p^{\prime}\over (2\pi)^{3}} \,v^{j}e^{-\beta_{0}(H_{\rm part}-\mu_{0})}=0.
\end{equation}
The normalized velocity vector of the ideal gas is  therefore $U^{\mu}=\delta^{\mu}_{ 0}/\sqrt{g_{00}}$,
which allows (\ref{420}) to be expressed in the perfect fluid form
\begin{eqnarray}
\langle T^{\mu\nu}_{\rm part}\rangle&=&e^{\beta_{0}(\mu_{0}-eA_{0})}
\Big\{U^{\mu}U^{\nu}(\overline{\rho}+\overline{P})-g^{\mu\nu}\overline{P}\Big\}\nonumber\\
&=&\Big\{U^{\mu}U^{\nu}(\rho_{\rm part}+P_{\rm part})-g^{\mu\nu}P_{\rm part}\Big\}.\label{430}
\end{eqnarray}
Because of the external potential $A_{0}({\bf x})$ the covariant divergence of $\langle T^{\mu\nu}_{\rm part}\rangle$
is not zero:
\begin{equation}\langle T^{\mu\nu}_{\rm part}\rangle_{;\mu}=g^{\nu j}{e\over T_{0}}(\partial_{j}A_{0}).
\end{equation}

\section{Discussion}
\paragraph{Low temperature example.} 
The spatial dependence of $\overline{P}$ comes entirely from
$g_{00}$ and so  $\overline{P}$ is constant on surfaces of constant $g_{00}$.
The full particle pressure $P_{\rm part}=e^{\beta_{0}(\mu_{0}-eA_{0})}\overline{P}$ 
has somewhat different isobaric surfaces if $A_{0}\neq 0$.
The result (\ref{391}) for $\overline{P}$ may be evaluated when $Z=m\sqrt{g_{00}}/T_{0}\gg 1$  using
the asymptotic behavior $K_{2}(Z)\to e^{-Z}\sqrt{\pi/2Z}$: 
\begin{equation}
P_{\rm part}={T_{0}\over\sqrt{g_{00}}}\Big({mT_{0}\over 2\pi\sqrt{g_{00}}}\Big)^{3/2}
e^{\beta_{0}(\mu_{0}-eA_{0}-m\sqrt{g_{00}}) }+\dots\label{500}.\end{equation}
The resulting thermodynamic functions (number density, entropy density, and energy density) are
displayed in Appendix C.
The gradient of the pressure is
\begin{equation}
{\partial_{j}P_{\rm part}\over P_{\rm part}}=-\beta_{0}\Big[m{\partial_{j}g_{00}\over 2g_{00}}+e\partial_{j}A_{0}\Big]
-{5\over 4} {\partial_{j}g_{00}\over(g_{00})^{3/2}}.
\end{equation}
For the Kerr-Newman metric in Boyer-Lindquist coordinates $t,r,\theta,\phi$ both $g_{00}$ and $A_{0}$ depend on 
$r$ and $\theta$. At large distances the $\theta$ dependence is negligible
\begin{eqnarray}
g_{00}&\approx& 1-{2GM\over r}\\
A_{0}&\approx &{Q\over r}
\end{eqnarray}
and the pressure gradient is radial:
\begin{equation}
{dP_{\rm part}\over dr}\approx n\Big( -{GMm\over r^{2}}+{eQ\over r^{2}}\Big)
\end{equation}
after using $\beta_{0}P_{\rm part}\approx n$ and neglecting the ${\cal O}(T_{0}/m)$ correction.  The first term in the parenthesis is the gravitational
force exerted on the particle by the central mass; the second term is the electrostatic force on
the particle.

\paragraph{Quantum statistics.}  The calculations  presented are for distinguishable
point particles obeying Boltzman statistics. If instead the particles are indistinguishable
and obey either Bose-Einstein or Fermi-Dirac statistics the particle pressure becomes
\begin{equation}
P_{\rm part}=-{T_{0}\over\xi \sqrt{g}}\int {d^{3}p\over (2\pi)^{3}}\ln\big[1-\xi e^{-\beta_{0}(H_{\rm part}-\mu_{0})}\big]
\end{equation}
where $\xi=1$ for Bose statistics and $\xi=-1$ for Fermi statistics. 
When expanded in a series
\begin{equation}
P_{\rm part}={T_{0}\over \xi\sqrt{g}}\sum_{b=1}^{\infty}{(\xi e^{\beta_{0}(\mu_{0}-eA_{0})})^{b}\over b}
\int {d^{3}p\over (2\pi)^{3}}e^{-b\beta_{0}\overline{H}}
\end{equation}
the integral  is the same as (\ref{304}) except that $\beta_{0}$ is replaced by $b \beta_{0}$. The
value of the integral may be read off from (\ref{391})
\begin{equation}
P_{\rm part}={m^{2}\over 2\pi^{2}\xi}\sum_{b=1}^{\infty}(\xi e^{\beta_{0}(\mu_{0}-eA_{0})})^{b}
\Big({T_{0}\over b\sqrt{g_{00}}}\Big)^{2}K_{2}\Big(mb{\sqrt{g_{00}}\over T_{0}}\Big)\label{505}
\end{equation}
and satisfies the Tolman-Ehrenfest rule except for the $eA_{0}$ dependence.
If the ensemble contains particles and antiparticles the total particle pressure requires adding
to (\ref{505}) another such series with $\mu_{0}$ replaced by $-\mu_{0}$.
A simple check of (\ref{505}) is that of  massless bosons with $\mu_{0}=A_{0}=0$ in which case
\begin{equation}
P_{\rm part}={\pi^{2}\over 90}\bigg({T_{0}\over\sqrt{g_{00}}}\bigg)^{4}.\end{equation}

         \begin{appendix}
          \section{Change to  Euclidean momenta} The calculation in Sec. III of the partition function requires integrating
$\exp[-\beta_{0}\overline{H}]$ with respect to the canonical momenta $dp_{1}dp_{2}dp_{3}$,
where
\begin{equation}
\overline{H}={1\over g^{00}}\bigg[\sqrt{g^{00}(m^{2}-c^{ij}p^{\prime}_{i}p^{\prime}_{j})}+g^{0j}p^{\prime}_{j}\bigg],\label{A10}
\end{equation}
 with $c^{ij}$ as defined in (\ref{236}), and $p_{j}^{\prime}=p_{j}+eA_{j}$. 
The spatial metric $g_{jk}$ may be expanded at any point $x^{i}$ in terms of three Euclidean frame vectors
\begin{equation}
g_{jk}=-\sum_{a=1}^{3} f_{(a)j}f_{(a)k},\label{A12}\end{equation}
which are the analogs of vierbeins in three-dimensional space. (The minus sign arises because the eigenvalues of $g_{jk}$
are negative.)
The  contravariant form of the frame vectors is
\begin{equation} 
f_{(a)}^{i}=-c^{ij}f_{(a)j}.\label{A15}\end{equation}
Equations  (\ref{A12}), and (\ref{A15}) imply
\begin{eqnarray}
\sum_{a=1}^{3}f_{(a)}^{i}f_{(a)j}&=&\delta^{i}_{\;\;j}\nonumber\\
f_{(a)}^{j}f_{(b)j}&=&\delta_{ab}\\
\sum_{a=1}^{3} f_{(a)}^{i}f_{(a)}^{j}&=&-c^{ij}.\nonumber
\end{eqnarray}
The quadratic term in $\overline{H}$ is
\begin{equation}
-c^{ij}p_{i}p_{j}=\sum_{a=1}^{3} (f^{i}_{(a)}p^{\prime}_{i})(f^{j}_{(a)}p^{\prime}_{j}),\label{A20}
\end{equation}
which suggests introducing  Euclidean momenta $k_{a}$ 
\begin{equation}
k_{a}=f^{j}_{(a)}\,p^{\prime}_{j}.\label{A25}\end{equation}
(This is not a canonical transformation: the defining Poisson bracket $\{x^{i},p_{j}\}=\delta^{i}_{j}$ implies
$\{x^{i}, k_{a}\}=f^{i}_{(a)}$.)
$\overline{H}$ is now
\begin{equation}\overline{H}={1\over g^{00}}\bigg[\sqrt{g^{00}(m^{2}+{\bf k}^{2})}+g^{0j}\sum_{a=1}^{3}f_{(a)j}k_{a}\bigg].
\end{equation}
Define a Euclidean vector
\begin{equation}
s_{a}={g^{0j}f_{(a)j} \over \sqrt{g^{00}}}\end{equation}
with length  
\begin{equation}
s^{2}=\sum_{a=1}^{3} [s_{a}]^{2}=g_{0j}g^{0j}=1-g_{00}g^{00}.\label{A30}
\end{equation}
The Hamiltonian  has the  simple form
\begin{equation}
\overline{H}={1\over \sqrt{g^{00}}}\Big[\sqrt{m^{2\!}+\!{\bf k}^{2}}+{\bf s}\cdot{\bf k}\Big].\label{A35}
\end{equation}
The inverse of  relation (\ref{A25}) is
\begin{equation}
p^{\prime}_{j}=\sum_{a=1}^{3} f_{(a)j}k_{a},
\end{equation}
 which gives for the Jacobian of the change from  $p^{\prime}_{j}$ to Euclidean $k_{a}$ 
\begin{eqnarray}d^{3}p^{\prime}&=&\sqrt{|{\rm det}(g_{ij}) }|\,d^{3}k=\sqrt{{g^{00}\over |{\rm det}(g^{\mu\nu})|}}d^{3}k\nonumber\\
&=&\sqrt{gg^{00}}\,d^{3}k\end{eqnarray}
where $g=|{\rm det}(g_{\mu\nu})|$.  These results  yield (\ref{312}).

          \section{$T^{\mu\nu}$ for particles and background field}
\subsection{Calculation of  $T^{\mu\nu}_{\rm part}$ from $H_{\rm part}$}
The energy-momentum tensor for the particles may be calculated using the Hilbert
variational principle:
\begin{eqnarray}
-{\sqrt{g}\over 2}T^{\mu\nu}_{\rm part}(x)&=&{\delta\over\delta g_{\mu\nu}(x)}\int dx^{\prime 0} L_{\rm part}(x')\nonumber\\
&=&\delta^{3}({\bf x}-{\bf x}^{\prime})\bigg[{\partial L_{\rm part}\over\partial g_{\mu\nu}}\bigg]_{\partial_{0}x^{i}, x^{i}}\label{B10}
\end{eqnarray}
where the metric variation is performed before the equations of motion are imposed on the particle
velocity and position.
The following steps depend  three facts: (i) $L_{\rm part}$ does not depend on the momentum $p_{j}$,
(ii) $L_{\rm part}=p_{j}\partial_{0}x^{j}-H_{\rm part}$ and $p_{j}\partial_{0}x^{j}$ does not depend on the metric,
(iii) $H_{\rm part}$ does not depend on the velocity $\partial_{0}x^{i}$:
\begin{eqnarray}
\bigg[{\partial L_{\rm part}\over\partial g_{\mu\nu}}\bigg]_{\partial_{0}x^{i},x^{i}}&=&
\bigg[{\partial L_{\rm part}\over\partial g_{\mu\nu}}\bigg]_{\partial_{0}x^{i},x^{i}, p_{j}}\label{B12}\\
&=&-\bigg[{\partial H_{\rm part}\over\partial g_{\mu\nu}}\bigg]_{\partial_{0}x^{i},x^{i}, p_{j}}\label{B13}\\\
&=&-\bigg[{\partial H_{\rm part}\over\partial g_{\mu\nu}}\bigg]_{x^{i}, p_{j}}.\label{B14}
\end{eqnarray}
Therefore
\begin{equation}
{\sqrt{g}\over 2}T^{\mu\nu}_{\rm part}(x)=\delta^{3}({\bf x}-{\bf x}^{\prime})\bigg[{\partial H_{\rm part}\over\partial g_{\mu\nu}}
\bigg]_{x^{i},p_{j}}.\end{equation}
The partition function for the particles 
\begin{equation}
\ln Z_{\rm part}=\int {d^{3}xd^{3}p\over (2\pi)^{3}}e^{-\beta_{0}(H_{\rm part}-\mu_{0})}\end{equation}
has a variational derivative 
\begin{equation}
{\delta \ln Z_{\rm part}\over\delta g_{\mu\nu}(x')}
=-\beta_{0}{\sqrt{g}\over 2}
\int {d^{3}pd^{3}x\over(2\pi)^{3}} T^{\mu\nu}_{\rm part}e^{-\beta_{0}(H_{\rm part}-\mu_{0})}\label{B20}
\end{equation}
This is the starting point of Sec. IV.

\paragraph*{Explicit form for $T^{\mu\nu}_{\rm part}$.}
It is not difficult to explicitly differentiate $H_{\rm part}$ with respect to $g_{\mu\nu}$
by considering the metric dependence of the covariant particle velocity $u_{\alpha}$.
From (\ref{255a})  $u_{j}$ is independent of the metric because $p_{j}$ and $A_{j}$ are.
The derivative of $g^{\alpha\beta}u_{\alpha}u_{\beta}=1$ with respect to $g^{\mu\nu}$
yields
\begin{equation}
{\partial u_{0}\over \partial g^{\mu\nu}}=-{u_{\mu}u_{\nu}\over 2u^{0}}.\end{equation}
This gives the derivative of  $H_{\rm part}=mu_{0}+eA_{0}$. 
A change from the contravariant metric to the covariant metric gives
\begin{equation}
\sqrt{g}\,T^{\mu\nu}_{\rm part}=\delta^{3}({\bf x}-{\bf x}^{\prime}) m{u^{\mu}u^{\nu}\over u^{0}}\label{B52}
\end{equation}
as expected.

\subsection{Calculation of  $T^{\mu\nu}_{\rm field}$  from ${\cal H}_{\rm field}$}
\subsubsection{The canonical ${\cal H}_{\rm field}$}
Because $g_{0j}\neq 0$ it is not trivial to compute ${\cal H}_{\rm field}$  in terms of the canonical momenta $\pi^{j}$ and the magnetic field $F_{jk}$.
The starting point is the Lagrange density 
\begin{equation}
{\cal L}_{\rm field}=-{\sqrt{g}\over 4}F_{\mu\nu}F_{\alpha\beta}g^{\mu\alpha}g^{\nu\beta},\end{equation}
which implies a canonical momentum 
\begin{equation}
\pi^{j}={\partial{\cal L}_{\rm field}\over\partial (\partial_{0}A_{j})}=-\sqrt{g}F^{0j}\end{equation}
and the Hamiltonian density 
\begin{eqnarray}
{\cal H}_{\rm field}&=&\pi^{j}\partial_{0}A_{j}+{\sqrt{g}\over 4}F_{\alpha\beta}F^{\alpha\beta}\nonumber\\
&=&\pi^{j}F_{0j}+{\sqrt{g}\over 4}F_{\alpha\beta}F^{\alpha\beta}+\pi^{j}\partial_{j}A_{0}\label{B60}
.\end{eqnarray}
To express this in terms of canonical variables requires the identity
\begin{equation}
g_{jk}\pi^{k}=-\sqrt{g}F^{0k}g_{kj}=-\sqrt{g}\,g^{0\mu}F_{\mu j}\label{B65}\end{equation}
which may be solved for $F_{0j}$:
\begin{equation}
F_{0j}={1\over g^{00}}\bigg[-{g_{jk}\pi^{k}\over\sqrt{g}}+F_{j\ell}g^{\ell 0}\bigg]
.\end{equation}
This allows the first term in (\ref{B60}) to be expressed in terms of momentum and covariant field strength.
For the second term in (\ref{B60}) use the quantity
\begin{equation}
c^{\alpha\mu}=g^{\alpha\mu}-{g^{\alpha 0}g^{\mu 0}\over g^{00}}\end{equation}
to obtain
\begin{equation}
F_{\alpha\beta}F^{\alpha\beta}=F_{\alpha\beta}c^{\alpha\mu}c^{\beta\nu}F_{\mu\nu}+2{F^{0\alpha}g_{\alpha\mu}
F^{0\mu}\over g^{00}}.\end{equation}
Because $c^{\alpha\mu}$ is zero if either $\alpha=0$ or $\mu=0$ and $F^{0\alpha}$ is zero if $\alpha=0$
the identity simplifies to 
\begin{equation}
F_{\alpha\beta}F^{\alpha\beta}=F_{jk}c^{j\ell}c^{km}F_{\ell m}+2{F^{0j}g_{jk}
F^{0k}\over g^{00}}.\label{B70}\end{equation}
This  allows the second term in (\ref{B60}) to be expressed as
\begin{equation}
{\sqrt{g}\over 4}F_{\alpha\beta}F^{\alpha\beta}={\pi^{j}\pi^{k}g_{jk}\over 2g^{00}\sqrt{g}}
+{\sqrt{g}\over 4}F_{jk}c^{j\ell}c^{km}F_{\ell m}.\label{B75}
\end{equation}
The resulting Hamiltonian density in terms of canonical variables is therefore
\begin{eqnarray}
{\cal H}_{\rm field}&=&-{\pi^{j}\pi^{k}g_{jk}\over 2g^{00}\sqrt{g}}+{\pi^{j}F_{jk}g^{k0}\over g^{00}}\nonumber\\
&+&{\sqrt{g}\over 4}F_{jk}F_{\ell m}c^{j\ell}c^{km}+\pi^{j}\partial_{j}A_{0}.\label{B80}
\end{eqnarray}
A simple check is  that  $\partial_{0}A{_j}=\partial{\cal H}_{\rm field}/\partial \pi_{j}$.

\subsubsection{The metric derivative of ${\cal H}_{\rm field}$}
The canonical momenta $\pi^{j}$ and the magnetic field $F_{jk}$ do not depend on the metric.
The metric dependence of ${\cal H}_{\rm field}$ due to $\sqrt{g}$ is easy to compute from (\ref{B80}) and gives
\begin{equation}
\bigg[{\partial\sqrt{g}\over\partial g^{\mu\nu}}\bigg]{\partial {\cal H}_{\rm field}\over \partial\sqrt{g}}
=-{\sqrt{g}\over 2}\Big({g_{\mu\nu}\over 4}F_{\alpha\beta}F^{\alpha\beta}\Big)
\end{equation}
after using (\ref{B75}) to convert canonical momentum back into fields.
To compute the metric derivative with $\sqrt{g}$ fixed
\begin{equation}
{\partial {\cal H}_{\rm field}\over\partial g^{\mu\nu}}\bigg|_{\sqrt{g}}
\end{equation}
is more difficult and must be calculated for the separate cases $\mu\nu=00,j0, jk$. 
After computing the metric derivatives it is necessary to use the relation
\begin{equation}
F^{0k}=g^{0\mu}F_{\mu j}c^{jk}\end{equation}
which follows from (\ref{B65}).  The final result is
\begin{equation}
{\partial{\cal H}_{\rm field}\over\partial g^{\mu\nu}}=-{\sqrt{g}\over 2}\bigg[-F_{\mu\alpha}F_{\nu\beta}g^{\alpha\beta}
+{g_{\mu\nu}\over 4}F_{\alpha\beta}F^{\alpha\beta}\bigg].\label{B85}
\end{equation}
A change from contravariant $g^{\mu\nu}$ to covariant $g_{\mu\nu}$ and the relation
 $\ln Z_{\rm field}=-\beta_{0}H_{\rm field}$ give
\begin{equation}
{\delta\ln Z_{\rm field}\over \delta g_{\mu\nu}}=-\beta_{0}{\sqrt{g}\over 2}T^{\mu\nu}_{\rm field}.\label{B90}
\end{equation}

         \section{Thermodynamics functions}
\subsection{Exact relations}
\paragraph{Thermodynamic quantities.}
The thermodynamic pressure due to the particles is
\begin{equation}
P_{\rm part}={T_{0}\over\sqrt{g}}\int {d^{3}p^{\prime}\over (2\pi)^{3}}
e^{-\beta_{0}(H_{\rm part}-\mu_{0})}.\label{C10}\end{equation}
Differentiation gives
\begin{equation}
\mu_{0}{\partial P_{\rm part}\over\partial \mu_{0}}+T_{0}{\partial P_{\rm part}\over\partial T_{0}}
=P_{\rm part}+\rho_{\rm part}+n{eA_{0}\over\sqrt{g_{00}}}.\label{C12}
\end{equation}
where
\begin{equation}
\rho_{\rm part}={1\over\sqrt{g}}\int {d^{3}p^{\prime}\over (2\pi)^{3}}\overline{H}\,
e^{-\beta_{0}(H_{\rm part}-\mu_{0})}.\label{C14}
\end{equation}
The first term in (\ref{C12}) is related to the number density:
\begin{equation}{n\over\sqrt{g_{00}}}={\partial P_{\rm part}\over\partial \mu_{0}}
=\beta_{0}P_{\rm part}.
\end{equation}
This  is  the ideal gas law
\begin{equation}
P_{\rm part}=n{T_{0}\over\sqrt{g_{00}}}\end{equation}
in local form.
The second term in (\ref{C12}) is related to the local entropy density
\begin{eqnarray}
s&=&\sqrt{g_{00}}\,{\partial P_{\rm part}\over\partial T_{0}}\nonumber\\
&=&{1\over T}\big[P_{\rm part}+\rho_{\rm part}+n{eA_{0}\over\sqrt{g_{00}}}-\mu n \big].\label{C28}
\end{eqnarray}

\subsection{Low temperature regime}
Using the approximation (\ref{500}) 
the number density is 
\begin{equation}
n({\bf x})=\Big({mT_{0}\over 2\pi\sqrt{g_{00}}}\Big)^{3/2}
e^{\beta_{0}(\mu_{0}-eA_{0}-m\sqrt{g_{00}}) }.\end{equation}
The entropy density $s({\bf x})$  can be computed from (\ref{C28}) and afterwards if the chemical potential is expressed in terms of the density the result is
\begin{equation}
s({\bf x})=n\bigg[{5\over 2}+\ln\bigg\{{1\over n}\bigg({mT_{0}\over 2\pi\sqrt{g_{00}}}\bigg)^{3/2}\bigg\}\bigg],\label{C40}
\end{equation}
which is the local form of the Sackur-Tetrode equation \cite{Pereira}. 
The energy density  of the particles is
\begin{equation}
    \rho_{\rm part}({\bf x})=n\bigg[m+{3\over 2}{T_{0}\over\sqrt{g_{00}}}\bigg].\label{C45}
\end{equation}

\end{appendix}

\end{document}